\documentstyle[11pt]{article}
\newcommand{\be}{\begin{equation}}
\newcommand{\ee}{\end{equation}}
\newcommand{\ba}{\begin{eqnarray}}
\newcommand{\ea}{\end{eqnarray}}
\newcommand{\bann}{\begin{eqnarray*}}
\newcommand{\eann}{\end{eqnarray*}}

\begin{document}
\hbadness=10000
\setcounter{page}{1}
\title{
\vspace{-2.5cm}
\hspace{-1.0cm}
{\huge \bf 
Higher Order Bose-Einstein Correlations test the Gaussian
Density Matrix and the Space-Time Approach}
}
\author{
N. Arbex$^1$\thanks{E. Mail: NELMARA@FMA.IF.USP.BR},
M. Pl\"umer$^2$\thanks{E. Mail: PLUEMER@SOULTEK.K.EUNET.DE} 
and
R.M. Weiner$^3,^2$\thanks{E. Mail: WEINER@MAILER.UNI-MARBURG.DE}
}

\date{$^1$ Physics Institut, S\~ao Paulo University, Brazil\\
$^2$ Physics Department, University of Marburg,
Germany \\
$^3$ LPTHE Orsay, Universit\'e Paris-Sud, France}
\maketitle

\begin{abstract}

A multiparticle system produced by a large number of independent sources
is described by a gaussian density matrix $\hat{W}$. 
All theoretical approaches to Bose-Einstein Correlations
$C_n$ in
high energy physics use this form for $\hat{W}$.
One of the most salient consequences of this form
is the fact that all higher order ($n>2$) moments of
the current distribution can be expressed in 
terms of the first two.
We test this property by comparing the data on $C_2(Q^2)$,
$C_3(Q^2)$ and $C_4(Q^2)$ from $\pi^+$p and $K^+$p reactions at $250$
$GeV/c$ with the predictions of a general quantum statistical
space-time approach.
Even a simplified version of such an approach involving
only 4 (instead of the total of 10) independent parameters
(proper-time, correlation-length, transverse radius of chaotic
source, chaoticity) can account for the data.
Previous attempts along these lines, which did not use the
space-time approach, met with difficulties.

\end{abstract} 

\begin{center}
{\small to appear in Physics Letters B (PLB 13397)}
\end{center}

\newpage

Bose-Einstein Correlations (BEC) are the basis of an experimental method
for the determination
of sizes and lifetimes of sources in particle and nuclear physics.
This knowledge is essential for an understanding of the 
dynamics of strong interactions.

A particularly important aspect of BEC is represented by
higher order correlations because of the prediction,
which follows from the gaussian density matrix assumption,
that all higher order moments of the current
distribution
can be described in terms
of the first two.
The gaussian form (in the coherent state
representation)
of the density matrix \footnote{The mathematical form of
the density matrix must not be confused with the form of
the correlator or of the space-time distribution of the
source, cf.below.} is a fundamental
assumption of BEC and follows from the central limit
theorem for a large number of independent sources,
which are expected to act in a high energy reaction.
Furthermore, higher order correlations provide important 
constraints on the space-time form of the sources, their 
dynamics (expansion) and chaoticity.

Experimentally correlations of three and more particles have been 
studied in the last years in \cite{17}-\cite{NA22} 
and more recently attempts have been made to analyze these correlations
in terms of simplified models \cite{Wei1}-\cite{Raz}
without clear space-time implications for the 
emitting source (sources).
Usually gaussian or exponential forms
for the correlation functions in momentum space
are postulated.
In contrast to this, 
the approach used in this work starts with the space-time
characteristics of an expanding source and 
the space-time form of the correlators
within the classical current formalism.
The dependence of $C_n$ on the four-momentum difference ($Q$)
follows after explicit
integration over space-time variables
\cite{bu}.

The aim of this investigation is to use the higher order
correlations NA22 data to test the validity of the gaussian
density matrix assumption, within the space-time approach to BEC.

\section{The General Formalism}

In quantum mechanics, a multiparticle production process is described
in terms of a density matrix $\hat{W}$ which characterizes the final
state of the system.
{}From the density matrix, all n-particle distributions can be
determined, and conversely, a measurement of these distributions
yields information about the density matrix of the multiparticle system.
The n-particle inclusive distribution is defined through 
the creation and annihilation operator 
$a_i^{\dag}({\bf k})$ and $a_i ({\bf k})$ of a particle
of momentum $k$  ($i$ labels internal
degrees of freedom): 

\begin{eqnarray}
\rho_n^{i_1...i_n}({\bf k_1},...,{\bf k_n})\equiv\mbox{\hskip8cm}\nonumber\\
\frac{1}{\sigma}\frac{d^n\sigma^{i_1...i_n}}{d\omega_1...d\omega_n}=(2\pi)^{3n}\prod^n_{j=1}2E_jTr(\hat{W}a^{\dag}_{i_j}({\bf k_1})...
a^{\dag}_{i_n}({\bf k_n})a_{i_n}({\bf k_n})...a_{i_1}({\bf k_1}))
\end{eqnarray}
where,
\begin{equation}
d\omega_i=\frac{d^3k_i}{(2\pi)^32E_i}
\end{equation}
is the invariant volume in momentum space.

The general n-particle correlation function is defined as
\begin{eqnarray}
C_n^{i_1...i_n}({\bf k_1},...,{\bf k_n})=
\frac{\rho_n^{i_1...i_n}({\bf k_1},...,{\bf k_n})}
{\rho_1^{i_1}({\bf k_1})...\rho_1^{i_n}({\bf k_n})}
= 1 + \frac{\bar{C}_n^{i_1...i_n}({\bf k_1},...,{\bf k_n})}
{\rho_1^{i_1}({\bf k_1})...\rho_1^{i_n}({\bf k_n})}. 
\end{eqnarray}

In order to determine the density matrix for a given reaction from first
principles, one would have to specify the initial state of the projectile
and target and then apply the $S$ matrix to this state.
In general this is not possible. One way to proceed is to parametrize 
$\hat{W}$ according to a reasonable phenomenological 
description of the system.
For this, one uses the external source (current) formalism.
In this approach particle sources are treated as external classical currents, 
and their fluctuations are described by a gaussian distribution.
This last choice can be justified by the fact that, 
if one has a superposition of
$N$ independent sources, 
the gaussian form follows from the central limit theorem 
in the limit of large $N$.

In the following we will also consider correlation functions
as functions of $Q^2$ :
\footnote{ As usual, 
$Q_n^2 = \sum_{i<j}^{n}$ $q_{ij}^2$; $q^2_{ij}= - (k_i - k_j)^2$;
$n \ge 2$ and $i,j=1,...,n$.}

\begin{equation}
C_n(Q^2)= 1 + \frac{I_n(Q^2)}{I^{\prime}_n(Q^2)}
\end{equation}

with ($i,j= 1, ..., n$),
 
\begin{equation}
I_{n}(Q^2)= \int dw_{i}...\int dw_{n}\mbox{\hskip0.1cm} 
\bar{C}_{n}^{i_1...i_n}(k_1,...,k_n)\mbox{\hskip0.1cm}
\delta [Q^2 + \sum^n_{i,j=1 (i<j)} (k_i - k_j)^2]
\end{equation}

\begin{equation}
I^{\prime}_{n}(Q^2)= \int dw_{i}...\int dw_{n}\mbox{\hskip0.1cm}
\rho^{i_1}_{1}(k_i)...\rho^{i_n}_{1}(k_n)\mbox{\hskip0.1cm}
\delta [Q^2 + \sum^n_{i,j=1 (i<j)} (k_i - k_j)^2] 
\end{equation}

Here we give a brief summary of the derivation
of Bose-Einstein correlation functions in the current formalism
\cite{gordo}.

The current can in general be written as the sum of a chaotic and a coherent
component, $J(x)=J_{chaotic}(x)+J_{coherent}(x)$.
The Gaussian current distribution is completely specified by its first two
moments:
$I(x)\equiv<J(x)>$=$J_{coherent}(x)$ and
the two-current correlator\\
$D(x,y)\equiv <J(x)J(y)>-<J(x)><J(y)>$.

$I(x)$ and $D(x,y)$ can be parametrized as 
$I(x) \propto f_c(x)$ and
$D(x,y) \propto f_{ch}(x) C(x-y) f_{ch}(y)$,
where $f_c(x)$ and $f_{ch}(x)$ are the space-time distributions of the
coherent and the chaotic components of the source. 
The primordial correlator $C(x-y)$ reflects intrinsic dynamical 
properties of the source.
It contains some characteristic
length (or time) scales $L$, so-called correlation lengths
(for a system in thermal equilibrium the correlation
length can be related to the inverse of the temperature).
 
In the general case of a partially coherent source, the single
inclusive distributions of pions can be expressed also as a sum of a
chaotic and coherent component $(i= +,-,0$ denotes the
charge):
\begin{equation}
\frac{1}{\sigma}\frac{d\sigma^i}{d\omega}=\left.
\frac{1}{\sigma}\frac{d\sigma^i}{d\omega}\right| _{chaotic}
+\left. \frac{1}{\sigma}\frac{d\sigma^i}{d\omega}\right| _{coherent}\\
\end{equation}
where,
\begin{equation}
\left.\frac{1}{\sigma}\frac{d\sigma^i}{d\omega}\right| _{chaotic} = 
D(k),
\end{equation}
\begin{equation}
\left. \frac{1}{\sigma}\frac{d\sigma^i}{d\omega}\right| _{coherent}=|I(k)|^2.
\end{equation}
$I(k)$ and $D(k)$ are the on-shell Fourier transforms
of $I(x)$ and $D(x,y)$, respectively.

In general, the chaoticity parameter will be momentum-dependent:
\begin {equation}
p(k)=\frac{D(k)}{D(k)+|I(k)|^2}.
\end{equation}

To write down the correlations functions in a concise form one
introduces the normalized current correlator:
\begin{equation}
d_{rs}=\frac{D(k_r,k_s)}{[D(k_r,k_r).D(k_s,k_s)]^{1/2}}
\end{equation}
where the indices $r,s$ label the particles.
Since $d(k_r,k_s)$ 
is in general a
complex number, one may prefer to express the correlation functions in
terms of the magnitudes and the phases of $d_{rs}$,
\begin{eqnarray}
T_{rs}&\equiv& T(k_r,k_s)=|d(k_r,k_s)|,\\
\phi_{rs}^{ch}&\equiv& \phi^{ch}(k_r,k_s)= \arg d(k_r,k_s)
\label{eq:phase}
\end{eqnarray}
and the phase of the coherent component
\begin{equation}
\phi_r^{co}\equiv \phi^{co}(k_r)= \arg I (k_r).
\end{equation}

With this notation we write the chaoticity parameter: $p_r \equiv p(k_r)$.

The normalized cumulant correlation functions $H_n$ for identical
charged particles are:

\begin{equation}
H_2^{++}(k_1,k_2)=2\sqrt{p_1(1-p1)p_2(1-p_2)}T_{12}
cos(\phi_{12}^{ch}-\phi_1^{co}+\phi_2^{co})+p_1p_2T_{12}^2\\
\end{equation}
\begin{eqnarray}
H_3^{+++}(k_1,k_2,k_3)=\frac{1}{3} p_1p_2p_3T_{12}T_{23}T_{31}
cos(\phi_{12}^{ch}+\phi_{23}^{ch}+\phi_{31}^{ch})\nonumber\\
+\sqrt{p_1(1-p1)p_2^2 p_3(1-p_3)}
T_{12}T_{23}cos(\phi_{12}^{ch}+\phi_{23}^{ch}
+\phi_3^{co}-\phi_1^{co})\nonumber\\
+\mbox{permutation of 1,2,3 \hskip3cm}
\end{eqnarray}
\begin{eqnarray}
H_4^{++++}(k_1,k_2,k_3,k_4)= \mbox{\hskip8cm} \nonumber\\
\frac{1}{4} p_1p_2p_3p_4T_{12}T_{23}T_{34}T_{41}
cos(\phi_{12}^{ch}+\phi_{23}^{ch}+\phi_{34}^{ch}+\phi_{41}^{ch})
\mbox{\hskip2cm}\nonumber\\
+\sqrt{p_1(1-p1)p_2^2 p_3^2 p_4(1-p_4)}T_{12}T_{23}T_{34}
cos(\phi_{12}^{ch}+\phi_{23}^{ch}+\phi_{34}^{ch}
+\phi_4^{co}-\phi_1^{co}) \mbox{\hskip0.5cm}\nonumber\\
+\mbox{permutation of 1,2,3,4 \hskip4.3cm}
\end{eqnarray}
            
The correlation functions $C_n$ (3) can be expressed in terms of $H_n$
by using the relation between the corresponding generating functionals
\cite{gordo}:
\begin{equation}
C_2^{i_1,i_2}( k_1,k_2 )= 1 + H_2^{i_1,i_2}(k_1,k_2)
\end{equation}
\begin{eqnarray}
C_3^{i_1,i_2,i_3}( k_1,k_2,k_3 )= \mbox{\hskip8cm}\nonumber\\
1 + H_2^{i_1,i_2}(k_1,k_2)+
H_2^{i_2,i_3}(k_2,k_3)
+H_2^{i_3,i_1}(k_3,k_1)+
H_3^{i_1,i_2,i_3}(k_1,k_2,k_3)
\end{eqnarray}
\begin{eqnarray}
C_4^{i_1,i_2,i_3,i_4}&(&k_1,k_2,k_3,k_4 )=\nonumber\\
 1 &+& H_2^{i_1,i_2}(k_1,k_2)+H_2^{i_1,i_3}(k_1,k_3)
  + H_2^{i_1,i_4}(k_1,k_4)+H_2^{i_2,i_3}(k_2,k_3)\nonumber\\
 &+& H_2^{i_2,i_4}(k_2,k_4)+H_2^{i_3,i_4}(k_3,k_4)\nonumber\\
 &+& H_2^{i_1,i_2}(k_1,k_2).H_2^{i_2,i_3}(k_2,k_3)
  + H_2^{i_1,i_3}(k_1,k_3).H_2^{i_2,i_4}(k_2,k_4)\nonumber\\
 &+& H_2^{i_1,i_4}(k_1,k_4).H_2^{i_2,i_3}(k_2,k_3)\nonumber\\
 &+& H_3^{i_1,i_2,i_3}(k_1,k_2,k_3)+H_3^{i_1,i_2,i_4}(k_1,k_2,k_4)
  + H_3^{i_1,i_3,i_4}(k_1,k_3,k_4)\nonumber\\
 &+& H_3^{i_2,i_3,i_4}(k_2,k_3,k_4)
  + H_4^{i_1,i_2,i_3,i_4}(k_1,k_2,k_3,k_4)
\end{eqnarray}

As one can see, all correlation functions
depend only on the functions
$T_{r,s}$, $\phi_{r,s}^{ch}$ and $\phi_r^{co}$ which 
will be specified in the next section for an 
expanding source. 

{}For the space-time description of an expanding source
is useful to define variables $\tau$, $\eta$ and $x_{\|}$:

\begin{eqnarray}
\tau=\sqrt{x_0^2-x_{\|}},\mbox{\hskip0.5cm}
\eta=\frac{1}{2}ln\frac{x_0+x_{\|}}{x_0-x_{\|}}
\end{eqnarray}
where $\tau$ is the proper-time, $x_{\|}$ the coordinate in the
direction of the collision axis and $\eta$ the space-time rapidity.
Here we will consider invariance under boosts of the coordinate frame 
in the longitudinal direction, which corresponds to the assumption
that the single inclusive 
distribution in rapidity is flat.

The space-time distribution of the chaotic and coherent source,
as well as the primordial correlator
\footnote{The parametrization for the primordial correlator,
in the case of an expanding source, has to take into account
that each source element 
is characterized not only by a correlation length but
also by a four-velocity. 
Effects of the geometry of the source are considered by
introducing the space-time distributions of 
the chaotic and the coherent
component, $f_{ch}(x)$ and $f_{co}(x)$ respectively.},  
are expressed in terms
of 10 parameters:
$\tau_{0,ch}$ and $\tau_{0,co}$ are the proper time
coordinates of the chaotic and the coherent source; $\delta \tau_{ch}$
and $\delta \tau_{co}$ their widths in proper-time; $R_{ch}$ and $R_{co}$
are the transverse radii; $L_{\bot}$ and $L_{\eta}$ are
the correlation lengths and $L_{\tau}$ is the correlation time.
The tenth one is the chaoticity parameter.

To reduce the number of parameters we assume now
that
the widths in proper-time of both sources are vanishing
($\delta\tau_{ch}=\delta\tau_{co}=0$); with this choice the
results do not depend on the correlation time $L_{\tau}$. 
The space-time
distributions of the chaotic and coherent sources are then
parametrized as:
\begin{eqnarray}
f_{ch}&\sim& \delta(\tau-\tau_{0,ch}) exp\left(-\frac{x_{\bot}^2}{R_{ch}^2}\right)\\
f_{co}&\sim& \delta(\tau-\tau_{0,co}) exp\left(-\frac{x_{\bot}^2}{R_{co}^2}\right)
\end{eqnarray}

Note that the $\eta$ dependence of $f_i(x)$ was neglected
in equations (22) and (23). This corresponds to a boost-invariant
ansatz of the source expansion.

The model contains now $7$ independent parameters:
$\tau_{0,ch}$, $\tau_{0,co}$, $R_{ch}$, $R_{co}$, 
$L_{\bot}$, $L_{\eta}$ and the chaoticity parameter $p_0$
\footnote{The relative contributions of the chaotic and coherent 
component are determined by fixing the value of the (momentum-dependent)
chaoticity parameter $p$ at same arbitrary scale, e.g.,
$p_0 \equiv p(k=0)$ (cf. Eq. (32)).}.

We will define, for convenience:
\begin{eqnarray}
b=\frac{\tau_{0,ch}^2}{2L_{\eta}^2},\mbox{\hskip0.5cm}
R_{L}^2=\frac{R_{ch}^2L_{\bot}^2}{R_{ch}^2+L_{\bot}^2}\mbox{\hskip0.5cm and}
\mbox{\hskip0.5cm}
\gamma_{12}=\frac{ \tau_{0,ch}(m_{1\bot}-m_{2\bot})}
{L_{\eta}^2m_{1\bot}m_{2\bot}}
\end{eqnarray}

The single inclusive distribution will be a sum of
a chaotic and a coherent term:
\begin{eqnarray}
E\frac{1}{\sigma}\frac{d^3\sigma}{d^3k}&=&[p_0 s_{ch}(k)+(1-p_0)s_{co}(k)]
E\left|\frac{1}{\sigma}\frac{d^3\sigma}{d^3k}\right|_{k=0}\\
s_{ch}(k)&=&\frac{m_{\pi}}{m_{\bot}}exp\left(-\frac{k_{\bot}^2R_{L}^2}{2}\right)\\
s_{co}(k)&=&\frac{m_{\pi}}{m_{\bot}}exp\left(-\frac{k_{\bot}^2R_{co}^2}{2}\right)
\end{eqnarray}
where $m_{\bot}$ is the transverse mass of the pions emitted and the momentum
dependence of chaoticity parameter takes the form $(r,s=1,2,3,4)$:
\begin{eqnarray}
p_r&=&p(k_r)=\frac{p_0}{A_r}\mbox{\hskip0.1cm},
\mbox{\hskip0.2cm}\mbox{\hskip0.1cm}A_r=p_0+(1-p_0)S_{rr}\\
S_{rs}&=&exp\left[-\frac{(k_{r\bot}^2+k_{s\bot}^2)(R_{co}^2-R_L^2)}{4}\right]
\end{eqnarray}

The magnitudes and phases of $T_{rs}$, $\phi_{rs}^{ch}$ and $\phi_r^{co}$
are:
\begin{eqnarray}
T_{rs}=(1+\gamma_{rs}^2)^{-1/4}exp\left[-\frac{b(y_r-y_s)^2}
{1+\gamma_{rs}^2}-\frac{(k_{r\bot}-k_{s\bot}^2)(R_{ch}^2-R_L^2)}{8}
\right]
\end{eqnarray}
\begin{eqnarray}
\phi_{rs}^{ch}&=&
\frac{b\gamma_{rs}}{1+\gamma_{rs}^2}
(y_r-y_s)^2-\tau_{0,ch}(m_{r\bot}-m_{s\bot})-
\frac{1}{2}\arctan \gamma_{rs}\\
\phi_{r}^{co}&=&-\tau_{0,co}m_{r\bot}
\end{eqnarray}

We proceed now to the the calculation
of the correlation functions.

\section{Calculations and Comparison with Experimental Data}

The procedure for calculations consists in:\\
a) The phase-space is generated 
by Monte Carlo routines which take into account the experimental
detection conditions mentioned in \cite{NA22,chNA22,infNA22} 
(defined by $|x_F|<0.5$) and references quoted there 
\footnote{We have also investigated the sensitivity of the fits to
the phase space varying the acceptance conditions. For this
purpose simulations for a different experimental window 
$-2<y<2$, $0.125GeV<k_t<1.5GeV$ were performed. We found that
in the range of $Q^2<0.5$$GeV^2$ the third and fourth order
correlation functions are more sensitive to the detections conditions
than the second order correlation function.}.
\\
b)The simulated produced pions are registered with a determined
$y, k_{\bot}$ and azimuth angle $\phi$ and from them one calculates
the correspondent $Q^2$ value.
The simulated particles 
are then selected in $Q^2$ bins, as described in the 
experimental papers quoted above.
\\
c)The calculation of the functions $T_{rs}$, $\phi_{rs}^{ch}$
and $\phi_{rs}^{co}$ defined for an expanding source model
is performed. From these results one is able to 
calculate the correlation function (and its integral for
each defined bin), 
as determined by equations (4)-(6) and (18)-(20).
\\
d)We use this procedure for second, third and fourth order
correlation calculations, $C_2(Q^2), C_3(Q^2)$ and $C_4(Q^2)$.

To simplify the calculations we reduce the number of 
free parameters using the criteria presented in \cite{bu}.

We assume at first that the chaotic and coherent
components have the same transverse momentum spectrum,
i.e., we choose $R_{co}=R_{L}$ 
(If $p_0 =1$, $R_{co}$ is obviously not present anymore  
in the formalism.).
$R_{L}$ is constrained by the
relation \footnote{$<k_t>$ is taken 
from experiment.}
$R_{L}=(\sqrt{\pi/2}$)$<k_t>^{-1}$.
We also take $\tau_{ch,0}=\tau_{co,0}\equiv \tau_0$.

We are thus left with only four parameters: $\tau_0, R_{ch},
L_{\eta}$ and $p_0$.

These parameters have been investigated in the intervals\footnote{The steps 
used in the calculation 
to adjust the parameters are
$\Delta R_{ch}= \Delta p_0 = \Delta \tau_0 = \Delta L_{\eta} = 0.1$.}:
$0.1\le p_0 \le 1.0$; $0.5 < R_{ch} < 3.5$;
$1.0 \le \tau_0 \le 2.5$ and 
$0.1 \le L_{\eta} \le 1.0$. 

We will use the $\chi^2/ndf$ ($ndf$: number of degrees of freedom)
method to evaluate the quality
of the fits. We are looking for a common fit
for all $C_i(Q^2)$ using one set of parameters 
($p_0$,$R_{ch}$,$\tau_0$,$L_{\eta}$).

The procedure to select this one set
 for an overall fit was:\\
1) Calculation of $\chi^2$ 
for each $C_i(Q^2)$ ($i=2,3,4$) function for all parameter sets
to delimit sectors in the parameters phase-space
to be investigated. 
The groups of parameters which gave 
$\chi^2/ndf$$>3$ for any calculated function were eliminated. 
\\
2) Search for regions of intersection in the parameters
space for an overall 
description of the data.
These regions were defined by minimizing $\chi^2/ndf$ for
all three functions simultaneously 
\footnote{The best fits do not necessarily present the lowest 
value for $\chi^2/ndf$ of {\it each} function,
since we look for regions in the parameter phase-space 
 which minimize
the $\chi^2/ndf$ for {\it all} three functions.}
and imposing
$\chi^2_{i}/ndf$$\le1.5$.

{}From 1) we concluded that 
there is a large number of parameter combinations which
permit acceptable fits for $C_2(Q^2)$. The situation is
quite different when one extends the search to
$C_3(Q^2)$ and $C_4(Q^2)$.

Table 1 shows the groups of parameters we would
like to comment on primarily. In all these groups $p_0=1$.
This table also contains the $\chi^2/ndf$ for each calculation.
Figure 1 shows the results for the
best overall fit
\footnote{We normalized the calculated correlation
functions by choosing $C_i(Q^2)=1$
for $Q^2=2.0$$GeV^2$ $(i=2,3,4)$.
The normalization factor is very close to one.}, corresponding 
to the parameters set $N2$
 \footnote{For completeness,
calculations have been done also for $C_n(Q)$ besides
$C_n(Q^{2})$. The $C_n(Q)$ functions did not lead to new conclusions
about the choice of parameters, they were in general 
agreement with the behaviour of $C_n(Q^2)$.}.

Some results for $p_0\ne 0$ should be commented.
Calculations using $p_0=0.3$ and $R_{ch}=1.7$$fm$
can produce good fits for $C_2(Q^2)$ ($\chi^2_2/ndf\le 1.3$)
in $0.2$$fm/c<L_\eta<0.7$$fm/c$  and in the entire $\tau_0$ interval, but
the minimum $\chi^2_i/ndf$ values for $C_3$ and $C_4$ are
around 2.0 in the (very small) region of intersection of parameters. 
The behaviour of the $C_2$ function does not
change even if the radius varies up to $2.5$$fm$, but then
there does not exist an intersection region of parameters 
(with $\chi^2_i/ndf$$\le1.5$) for $C_3$ and $C_4$.
The calculations using $p_0=0.7$ presented almost the same behaviour.  

We found a clear tendency towards a common value of $L_{\eta}=0.3$$fm/c$.
The prefered range of
the radius is $1.7$$fm<R_{ch}<2.2$$fm$.
The range  of $\tau_0$ seems to be connected with that of 
the radius. One gets the best fits for $R_{ch}=1.7$$fm$ with 
$\tau_0 = 1.1$$fm/c$ 
and for $R_{ch}=2.2$$fm$ with $\tau_0=1.5$$fm/c$  (see Table 1)
\footnote{ We have also calculated the statistical errors associated
with our fits. This was done by fixing, at a given run, 3 parameters
out of the total 4 and searching until an increase of $\chi^2_i$
value by one unity was obtained. The results for the set $N2$
($p_0$,$R_{ch}$,$L_{\eta}$,$\tau_0$) are:
$(0.6-1.0, 2.2, 0.3, 1.5)$; $(1.0, 1.8-3.4, 0.3, 1.5)$; 
$(1.0, 2.2, 0.2-0.4, 1.5)$ and $(1.0, 2.2, 0.3, 1.2-2.3)$.}.

As already mentioned, the number of parameter sets
which provide an acceptable fit for $C_2(Q^2)$ 
is much bigger than the number of parameter sets which provide
an acceptable general fit. This is exemplified in Figure 2.
This figure should also also clarify 
the method used in this work, which differs from the one used in 
\cite{pprUA1}, where the parameters obtained from a fit
 of $C_2$ data were used to ``predict"  $C_3$.
  
For the results in Figure 2 we
used: $p_0=0.7$, $R_{ch}=1.8$$fm$, $\tau_0=2.2$$fm/c$ and $L_\eta=0.5$$fm/c$.
The corresponding  $\chi^2/ndf$ values for $C_2$, $C_3$
and $C_4$ functions are: $1.30$, $7.05$, $1.83$.

\section{Discussion and Conclusions}

Our main conclusion is that in the frame
of a model based on quantum statistical principles and
a space-time picture of an expanding source,
a general description of the higher order correlations
data with the same parameters
\footnote{Given the simplifications made above resulting in
the reduction of the number
of parameters from a minimum of 10 to only 4, the
quantitative estimates obtained above for the radius,
correlation
length, life-time and chaoticity, although reasonable 
from the physics point of view,
must not be
overemphasized, the more so that the data on which these
estimates are based, have large errors.},
as those appearing in the first two correlation
orders, is possible. The assumption 
of a gaussian form for the density matrix is 
consistent with the data. 

A similar but weaker
conclusion was reached in \cite{Raz} . In \cite{Raz} it  
had been shown that the gaussian form of the density matrix
is robust enough to resist attempts of 
falsification \cite{UA1-old} \footnote{For a more detailed
discussion of this issue cf. also the comments preceding the
reprinted paper by Neumeister et al. in \cite{Wei}.}. 

On the
other hand in \cite{UA1-old} and \cite{Raz} the
momentum-space approach was used, the
currents were assumed to be real and no simultaneous fit of
all correlation functions has been performed. 
Similar caveats apply to 
\cite{NA22} where an analysis of the same NA22 data as
those investigated in the present paper 
was presented and where only ``marginal" agreement with
the simplified quantum optical model of 
\cite{Wei1,Biya-old} 
was found.
The fact that in the present
work we could fit the same data supports the space-time approach
and the necessity of a simultaneous fit.

Besides the fact mentioned already that
correlators in momentum-space
are associated with a four dimensional
``radius" which has no clear physical interpretation, the
space-time approach and the momentum-space approach 
differ also
from a purely mathematical
point of view.
In \cite{Wei1,Biya-old,Raz} the $Q$ dependence
is postulated from the beginning while in
the present paper, as in \cite{bu}, 
it results from a complex process of integration.

In \cite{pprUA1} the same momentum-space model (\cite{Wei1,Biya-old,Raz}) 
as that used in \cite{NA22}
was applied and compared with UA1-collaboration data. This time
a new technique for estimating the correlation data
was used but only second and third order correlations were considered.
At first the best fit parameters set for the second
order correlation data was established and then used to
{\em predict} the third
order correlation function, which was found to disagree
with the measured one. 
Given the insensitivity of of the fit
parameters found by us on the $C_2$ function, 
this result is not surprising and the
fitting procedures used in \cite{pprUA1} (as well as in
in \cite{NA22}) have to be qualified. 
Furthermore the general reservations expressed above
about the momentum-space approach apply here again.

One may put the question why it is necessary
to invoke higher order correlations
to constrain the source parameters whereas
the presence of a gaussian density matrix implies that the 
first and second moment of the current
distribution, $I(k)$ and $D(k_1,k_2)$,
determine all higher order correlations
\footnote{There is of course the fact that the phases $\phi$ of 
eq.(\ref{eq:phase})
enter in different combinations in the various correlation
functions. This is true both in the momentum-space approach
\cite{Wei1,Biya-old,Raz},
 as well as in the space-time approach. 
In the application to
data  \cite{UA1-old} and 
\cite{NA22} of the quantum optical approach this complication
  was ''circumvented" by
assuming that the fields are real.}.

In the applications of the space-time approach \cite{gordo} 
 one has 
to take into account that 
due to the limited
statistics, most experimental BEC data including those
analyzed in the present paper 
give the correlations in one single variable, $Q$.
On the other hand, the correlation function $C_2$, for instance,
depends on 6 variables: the three momenta of the pions of the
pair, $\vec{k}_1$ and $\vec{k}_2$
\footnote{In the case of symmetries of the source geometry, 
the number of independent variables is slightly reduced:
e.g., for the boost-invariant
azimuthally symmetric expanding source discussed
here, there are four independent variables.}.
Thus a large amount of information is lost due to the integration
that projects out the $Q$ dependence (cf. (5) and (6)).
 This is a fortiori true for higher order correlations.
Therefore phenomenologically
 higher order correlation data can play
an important role in constraining the source parameters.

On the other hand 
the fact that we were able to account for the data
even within this simplified approach, 
proves that
there are still many degrees-of-freedom not used at
the present level of theory/data comparison 
so that the challenge of disproving the gaussian density 
matrix
will remain a hard task for a long time ahead.

\vspace{2cm}

We are indebted to W. Kittel for an instructive 
correspondence and to B. Schlei, U. Ornik and O.A.M. Helene
for important discussions. N.A. was partially supported
by CNPq (Brazil) and by FAPESP (98/01446-0).

\newpage

{\Huge \bf Figure Captions:}

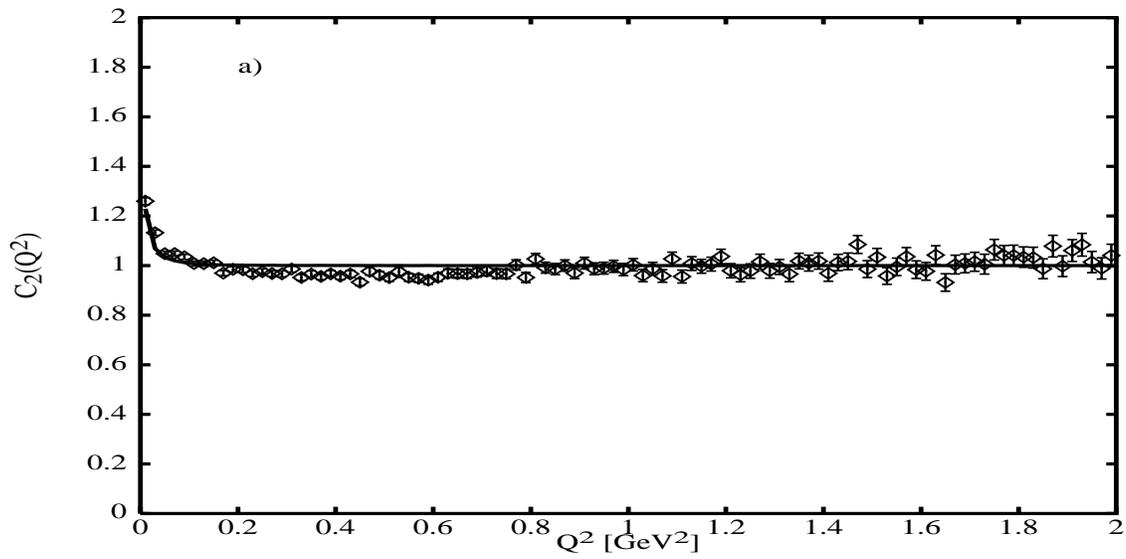
{\bf Figure 1:} The two-, three- and four-particle
correlation functions in $Q^2$ 
calculated using the space-time model defined in the text
 (continous lines), for parameters $N2$ (see Table 1)
compared with experimental data from \cite{NA22}.

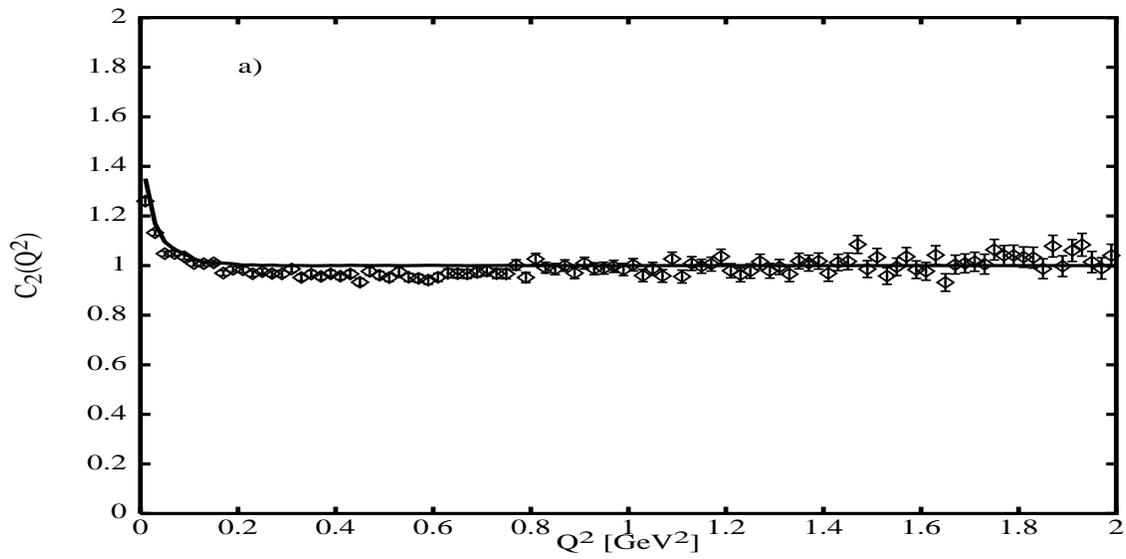
{\bf Figure 2:} The same calculation as in Figure 1 
using the other
set of parameters as defined in the text. The figure 
shows that parameters
which produce acceptable fits 
for $C_2(Q^2)$ do not necessarily
 give good overall fits. 

\vspace{1cm}

{\Huge \bf Table Captions:}

{\bf Table 1:} $\chi^2/ndf$ (last three columns)
for each correlation function for the parameter
sets $N1$,$N2$ and $N3$.  

%%%%%%%%%%%%%%%%%%%%%%%%%%%%%%%%%%%%%%%%%%%%%%
\newpage

\begin{center}
{\large \bf Table 1}
\end{center}
\begin{center}
\begin{tabular}{|c||c|c|c|c||c|c|c|}\hline
sets&$p_0$&$R_{ch} [fm]$&$L_{\eta} [fm/c]$&$\tau_{0} [fm/c]$&$C_2(Q^2)$&$C_3(Q^2)$&$C_4(Q^2)$\\
\hline
N1&1.0&1.7&0.2&1.1&1.35&1.23&1.49\\ \hline
N2&1.0&2.2&0.3&1.5&1.28&1.11&1.43\\ \hline
N3&1.0&2.5&0.3&1.3&1.31&1.12&1.41\\ \hline
\end{tabular}
\end{center}

%%%%%%%%%%%%%%%%%%%%%%%%%%%%%%%%%%%%%%%%%%%%%%%

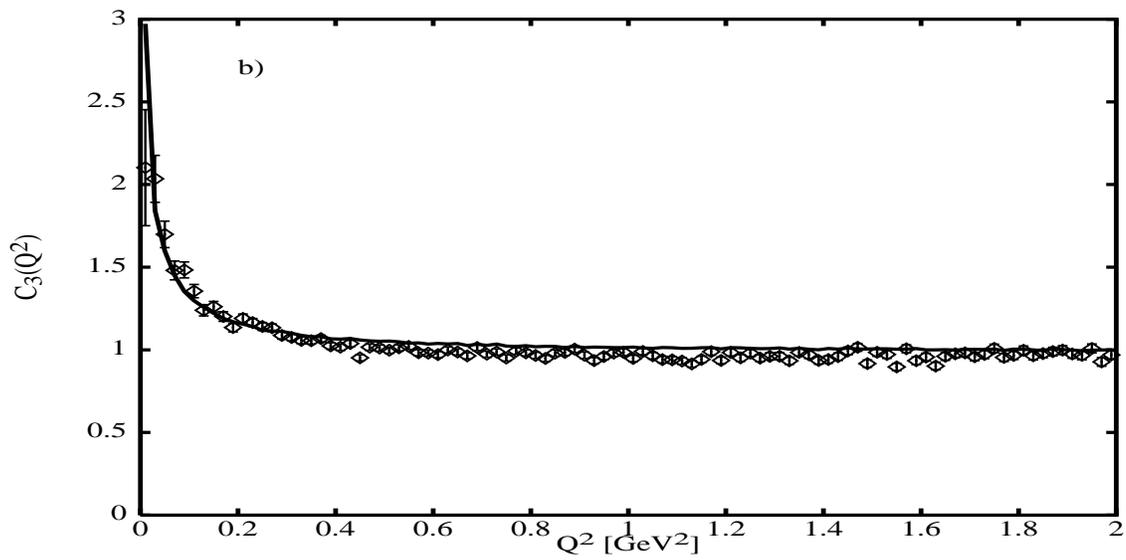
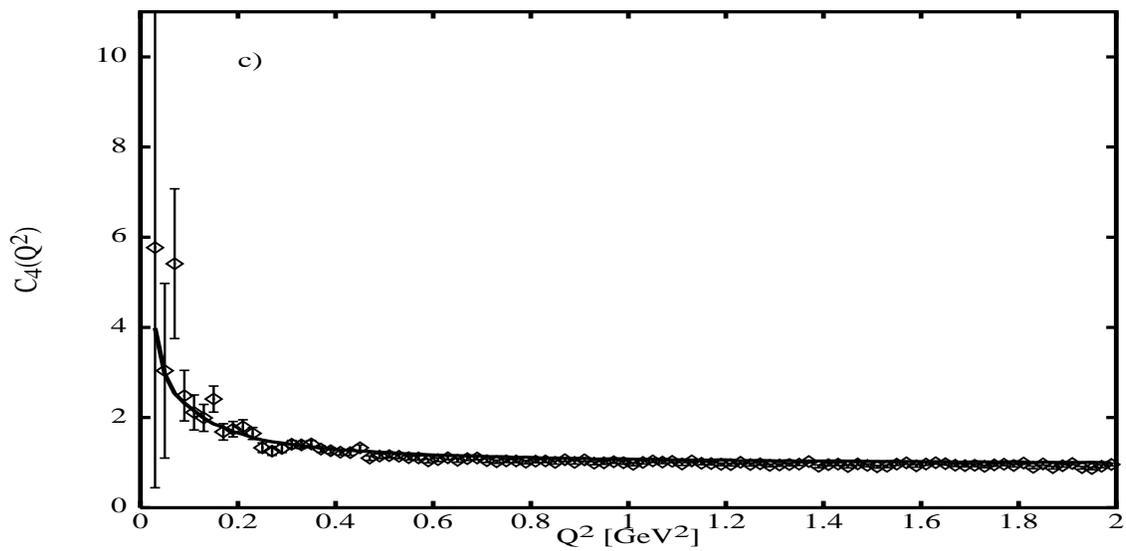
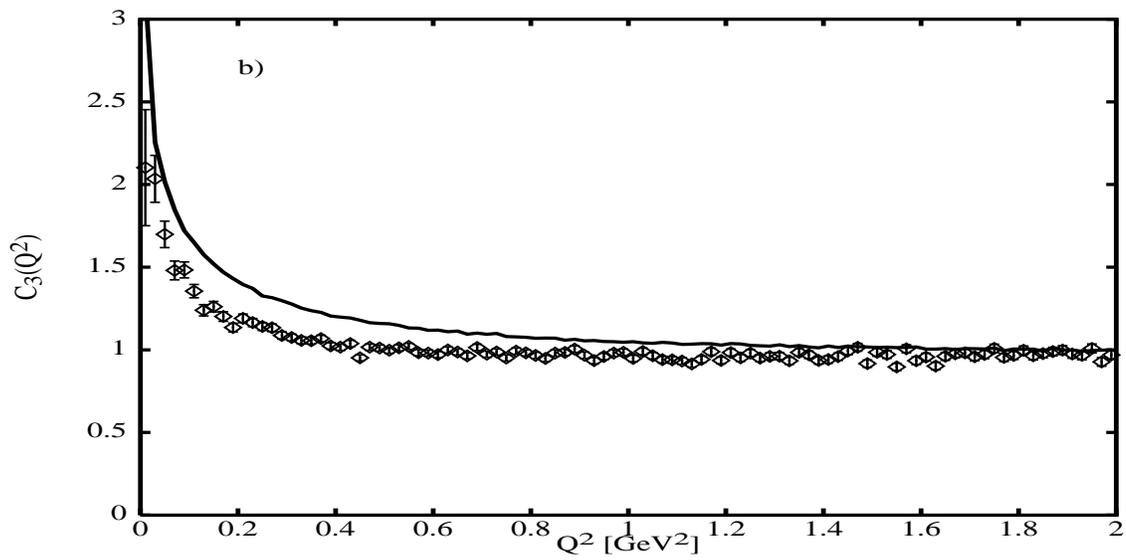
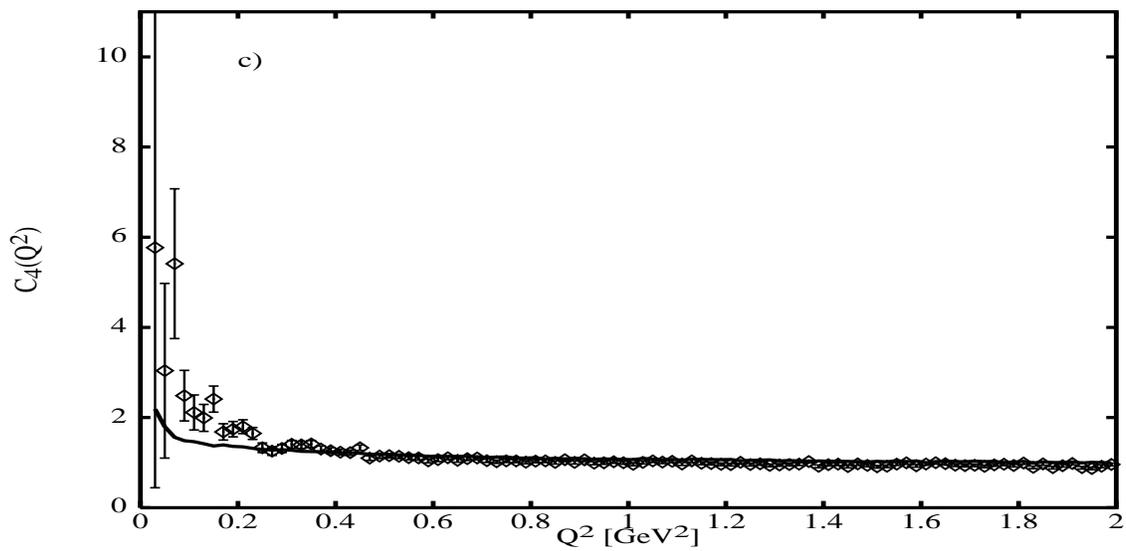
\end{document}